\documentclass[a4paper,fleqn,usenatbib]{mnras}

\usepackage[T1]{fontenc}
\usepackage{ae,aecompl}
\usepackage{hyperref}

\usepackage{graphicx}    % Including figure files
\usepackage{amsmath}    % Advanced maths commands
\usepackage{amssymb}    % Extra maths symbols

\newcommand{\source}{4FGL J0822.8--4207\,}

\title[Probing the origin of 4FGL J0822.8--4207]{Probing the origin of 4FGL J0822.8--4207: cosmic ray illumination from the SNR Puppis A and the Herbig-Haro object HH219}

\author[Araya et al.]{
M.~Araya,$^{1}$\thanks{E-mail: miguel.araya@ucr.ac.cr}
L.~Guti\'errez,$^{1}$
S.~Kerby$^{2}$
\\
$^{1}$Escuela de F\'isica, Universidad de Costa Rica, Montes de Oca, San Jos\'e, Costa Rica, 11501-2060\\
$^{2}$Department of Astronomy and Astrophysics,
 Pennsylvania State University, University Park, PA 16802, USA}

\date{Accepted 2021 November 15. Received 2021 October 22; in original form 2021 September 7}

\pubyear{2021}

\begin{document}
\label{firstpage}
\pagerange{\pageref{firstpage}--\pageref{lastpage}}
\maketitle

\begin{abstract}
\source is a point source found in the 4FGL-DR2 catalog by the gamma-ray observatory \emph{Fermi}-LAT and has no known association. We carry out X-ray observations of \source to help understand its nature. We explore two scenarios for the origin of \source. In the first case we study the possibility that cosmic rays from the supernova remnant (SNR) Puppis A, seen nearby in the sky, reach the dense gas at the location of the source and produce the gamma-rays through inelastic proton-proton collisions. We apply a standard model for particle diffusion in the interstellar medium and derive the required physical parameters. We find that this scenario for the gamma rays is possible if the gas is located at a distance that is not higher than $\sim 40$ pc from Puppis A, unless the SNR is older than 7 kyr or the diffusion coefficient is higher than typical Galactic values, and relatively low-energy cosmic rays are currently escaping from the SNR. In the second scenario, we consider the protostellar jet HH219 as the origin of the GeV source and find the very interesting possibility that particles could be accelerated up to energies of at least several TeV in HH219. This would make this system the first known of its kind to produce gamma-ray emission extending up to hundreds of GeV without any apparent cutoff and an excellent laboratory to study the process of particle acceleration.
\end{abstract}


\section{Introduction} \label{sec:intro}
Supernova remnants (SNRs) are known sources of gamma rays with energies from GeV to TeV, produced by relativistic particles that gain energy from the initial blast wave \citep[e.g.,][]{2015A&A...580A..74A,2016ApJS..224....8A} through a mechanism based on the idea of diffusive shock acceleration (DSA). It has been confirmed that in some SNRs the gamma rays result from inelastic collisions of protons and atomic nuclei with other ambient protons which result in the production and decay of neutral and charged pions \citep{2013Sci...339..807A}. Studying this hadronic emission is important as it provides information on the parent particle properties, which is relevant to the question of the origin of Galactic cosmic rays (CRs) up to energies of $\sim 1$ PeV \citep[for a review, see][]{2013A&ARv..21...70B}. Other sources/systems such as powerful star-forming regions could also be responsible for a part or the bulk of the particles \citep{2019NatAs...3..561A}.

No conclusive evidence for hadronic gamma rays caused by PeV particles has been obtained from SNR observations. A possibility is that SNRs do not accelerate particles to PeV energies. Another possibility is that even if PeV particles are produced in SNRs, their short residence time in these systems would make an observation difficult. In the theory of DSA, CRs are scattered by self-generated magnetic turbulence, but the highest-energy particles in the shock upstream region will experience a lack of turbulence and thus escape \citep{2003A&A...403....1P}. However, escaping particles could produce detectable signals of gamma rays and neutrinos after reaching interstellar clouds which can provide dense target material for hadronic interactions, thus serving as indirect probes of the particle acceleration in SNRs in the past \citep{1996A&A...309..917A,Gabici_2007,2009MNRAS.396.1629G,2011MNRAS.410.1577O,2021MNRAS.503.3522M}.

Studying the gamma-ray emission from clouds ``illuminated'' by CRs from a source located within several tens to $\sim 100$ pc \citep{1996A&A...309..917A} can provide information about the way particles diffuse through the interstellar medium (ISM), and reach and/or penetrate the cloud of gas. Measuring the spectral shape of the cloud emission can help constrain, according to theoretical models, the energy dependence of the diffusion coefficient in the propagation region \citep{2011MNRAS.410.1577O}. The known systems where this process is believed to occur are W28  \citep{2008A&A...481..401A,2010MNRAS.409L..35L,2014ApJ...786..145H} and perhaps also the $\gamma-$Cygni \citep{2020arXiv201015854M} and W44 \citep{2012ApJ...749L..35U,2020ApJ...896L..23P} SNRs.

In this work, we explore the origin of the GeV source \source, discovered by the \emph{Fermi} Large Area Telescope (LAT) and seen near the SNR Puppis A (G260.4--3.4) in the sky \citep{2020arXiv200511208B}. This dim gamma-ray source is point-like for the LAT and has no known association. It is seen $\sim0.9\degr$ from the geometrical centre of Puppis A, which is itself also a gamma-ray source \citep{2012ApJ...756....5L,2017ApJ...843...90X}. The gamma-ray emission of \source is not considered variable in the 4FGL catalog. As shown by \cite{2020ApJS..247...33A}, the variability index obtained from the one-year light curves of the sources that are not likely variable follows a $\chi^2$ distribution with $N-1$ degrees of freedom, where $N$ is the number of yearly intervals used. Variability is considered probable when the index of a source is above the 99\%-confidence in a $\chi^2(N-1)$ distribution (i.e., a $p-$value of 0.01). The variability index reported for \source is 20.97, slightly below the threshold value of 21.67 for $N=10$, the number of years of observation for the construction of the 4FGL-DR2 catalog  \citep{2020ApJS..247...33A,2020arXiv200511208B}. However, given the low significance of the GeV source, variability could be confirmed in the future with more observations.

The location of \source is consistent with the location of the star forming region G259.7592--02.8378, with a systemic velocity $v_{lsr}=10.7$ km s$^{-1}$, and is also very near G259.7743--02.8799, with $v_{lsr}=10.5$ km s$^{-1}$ \citep{2007A&A...474..891U}, which are consistent with the velocity derived for Puppis A in recent works \citep[e.g., $10.0\pm 2.5$ km s$^{-1}$,][]{2017MNRAS.464.3029R}. \cite{2007A&A...476.1019M} show three massive stellar objects in the region. At least a dense clump of gas with a diameter of 0.29 pc and a mass of 420 M$_\odot$ (assuming a distance of 1.7 kpc to the cloud) is known to exist at the location of \source \citep{2005A&A...432..921F}.

A protostellar jet known as HH219 is found at a location consistent with the positional uncertainty region of the GeV source \citep{1991PASP..103...79G}, and it is found approximately at the coordinates (J2000) RA$=08^h 22^m 54^s$, Dec$=-42\degr 08\arcmin 47\arcsec$ (RA$=125.725\degr$, Dec$=-42.146\degr$). The ejection processes resulting from the formation of massive stars can produce jets whose termination shocks and ambient conditions could accelerate particles to very high energies \citep{2007A&A...476.1289A,2010A&A...511A...8B}. For example, the typical life time for the jets is estimated to be a few kyr, while the timescales for particle acceleration could be of the order of tens of years \citep{2010A&A...511A...8B}. Furthermore, the presence of relativistic electrons has been confirmed in some protostellar jets from the detection of non-thermal radio emission \citep[e.g.,][]{2003ApJ...587..739G}.

Puppis A is a well-known SNR that has been studied across the electromagnetic spectrum. Despite the numerous observations in the region of this SNR, fundamental properties such as its distance are not known with certainty. A distance of $\sim2.2$ kpc was proposed by \cite{1995AJ....110..318R} assuming that an interaction between neutral hydrogen occurs in the eastern edge of the SNR, where an X-ray enhancement is detected. Age estimates ranging from 4 to 10 kyr have also been derived \citep{1985ApJ...299..981W,2003AAS...203.3912B,2012ApJ...755..141B}. Several neutral and molecular hydrogen clouds are seen in the region of the sky around Puppis A and some were believed to interact with the remnant, particularly towards its eastern side and at a systemic velocity of 16 km s$^{-1}$ \citep{1988A&AS...75..363D}. This proposed interaction could explain the shape of the eastern part of the remnant, although not the overall box-like appearance of the shell. This systemic velocity was found to be consistent with a kinematic distance of 2.2 kpc and therefore previous distance estimates \citep{1988A&AS...75..363D}. However, more recent studies have questioned the existence of an interaction with the large eastern cloud and the SNR, given the lack of any direct evidence, and have obtained instead a considerably lower distance to the source of 1.3 kpc \citep{2000MNRAS.317..421W}. A detailed H I study of Puppis A has determined a systemic velocity between 8 and 12 km s$^{-1}$ for the remnant, and a corresponding kinematic distance of $1.3\pm 0.3$ kpc \citep{2017MNRAS.464.3029R}. At this lower distance, no clouds are observed to the east that could account for the flat morphology of the shell, but radio polarization observations indicate that the overall shape of the SNR shell can be explained by the effect of the magnetic field into which it is expanding \citep{2018MNRAS.477.2087R}. As mentioned earlier, this recent systemic velocity for Puppis A of $10.0\pm 2.5$ km s$^{-1}$ is consistent with those reported for the star forming regions G259.7592--02.8378 and G259.7743--02.8799, which could mean that the dense clouds of gas seen in these regions are physically close to Puppis A.

No possible sources of high-energy emission are known at the location of \source, such as non-thermal radio or X-ray sources, except for the protostellar jet HH219. Given all these properties, our aim is to explore two possible mechanisms for the origin of the gamma rays. The first one is an origin resulting from cosmic ray illumination from Puppis A. Based on the presence of dense gas that can serve as a target for hadronic interactions resulting in high-energy gamma rays having a systemic velocity that is consistent with that recently derived for the SNR Puppis A, seen nearby in the sky, we explain the origin of \source in the context of cosmic ray illumination. The second scenario is related to particle acceleration in the protostellar jet HH219. The properties of the gamma-ray emission are consistent with those predicted by theoretical expectations for these objects \citep{2007A&A...476.1289A}. We also analyzed X-ray observations taken by the \textit{Swift} X-ray Telescope at the location of \source.

In section \ref{LAT} we present our analysis of LAT data to obtain the source spectrum and morphology. In section \ref{XRT} we present the X-ray observations of HH219 in the $0.3-10.0$ keV energy range. In section \ref{model} we apply the model for cosmic ray diffusion from Puppis A to the location of \source in order to explain the spectrum of the source. We also compare the properties that have been predicted for the high-energy emission expected from protostellar jets with those of the observed source. In section \ref{summary} we summarize the results.

%%%%%Articulo del modelo principal: 2011MNRAS.410.1577O, Ohira et al. 2011
%%% ejemplo del calculo original del espectro para r=20 pc, d=1kpc, ... http://articles.adsabs.harvard.edu//full/1996A%26A...309..917A/0000917.000.html -> comparar
%%%En la posicion de la fuente 4FGL puede haber protostellar jets, que se han visto en gammas, aunque con indice suave, https://arxiv.org/abs/1908.10994
%%%modelo teorico para HH objects: https://www.aanda.org/articles/aa/full_html/2010/03/aa13488-09/aa13488-09.html, https://ui.adsabs.harvard.edu/abs/2010A%26A...511A...8B/abstract
%%%YSOs as gamma-ray sources: https://www.aanda.org/articles/aa/full_html/2011/06/aa16580-11/aa16580-11.html
%%%emission could arise from particles accelerated in the colliding winds of massive stars or in the shocks produced by the outflows of protostars: https://arxiv.org/pdf/2107.12849.pdf

\section{LAT data} \label{LAT}
We analyzed data collected by \emph{Fermi}-LAT from the beginning of the mission, August 2008, to April 2021. Only events with energies above 1 GeV were included in the analysis in order to avoid the effects of the relatively poor point spread function (PSF) of the LAT at lower energies, and particularly to lower the influence of the bright Vela pulsar, located $\sim3 \degr$ from the SNR Puppis A in the sky \citep{Hewitt_2012}. We included events reconstructed in the region of interest (ROI), within $15\degr$ of the location of the source \source. We used the publicly available software {\tt fermitools}\footnote{See https://fermi.gsfc.nasa.gov/ssc/data/analysis/software/} version 2.0.8 and the open-source {\tt PYTHON} package {\tt fermipy}\footnote{See https://fermipy.readthedocs.io/en/latest/} version 1.0.1 to analyze the data. {\tt PASS8} data are filtered applying  recommended quality cuts such as {\tt DATA\_QUAL>0}, {\tt LAT\_CONFIG==1} and a zenith angle cut for events at less than $90\degr$ to avoid contamination from the Eath's limb. We combined back and front-converted events in the {\tt SOURCE} class with {\tt evtype=3} and {\tt evclass=128}. The background is modeled with the cataloged 4FGL-DR2 sources found within $20\degr$ of the ROI centre \citep{2020ApJS..247...33A,2020arXiv200511208B} including the Galactic diffuse emission and the isotropic component recommended for the data set and cuts, given by the files {\tt gll\_iem\_v07.fits} and {\tt iso\_P8R3\_SOURCE\_V3\_v1.txt}, respectively. We applied the energy dispersion as recommended by the LAT team\footnote{See https://fermi.gsfc.nasa.gov/ssc/data/analysis/\\documentation/Pass8\_edisp\_usage.html}. We originally excluded the source \source from the model in order to carry out a detailed study of the source morphology, location and spectral shape.

We used the binned maximum likelihood method \citep{1996ApJ...461..396M} to find the best-fit morphological and spectral parameters of the sources. For a new source with one additional parameter with respect to the null hypothesis (that is the background model without including the source), the method also allows for a calculation of its detection significance as the square root of the test statistic, defined as TS $=-2\times \ln (\mathcal{L}_0/\mathcal{L})$, with $\mathcal{L}_0$ and $\mathcal{L}$ the values of the maximum likelihoods for the null hypothesis and for a model with the additional source, respectively.

In order to take advantage of the improved PSF at higher energies, we carried out a morphological study using events with energies above 5 GeV only. We started by freeing the spectral normalizations of the sources located within $10\degr$ of the ROI centre and freeing the spectral shape parameters of the sources located within $5\degr$ of the ROI centre. We also freed the normalizations of the Galactic diffuse emission and the isotropic component. We performed a fit and found the parameter values, then removed the source \source from the model. We ran the tool {\tt find\_sources} to look for new maxima separated by more than $0.3\degr$ in the TS map of the null hypothesis. We freed the spectral parameters of the new sources (described by power-law functions) and repeated the fit. In the centre of the ROI the algorithm found a source consistent with the cataloged position of \source with TS$=38.9$. We ran a localization optimization algorithm (using the tool {\tt localize}) on this source and obtained the coordinates of its location with their 68\%-confidence level positional uncertainty, RA (J2000) $= 125.71 \pm 0.03 \degr$, Dec (J2000) $=-42.13\pm 0.03 \degr$. Since the cataloged position of \source is given by RA $=125.72\pm 0.05 \degr$, Dec $=-42.13 \pm 0.05 \degr$ \citep{2020ApJS..247...33A,2020arXiv200511208B} we concluded that the source found corresponds to \source. We also found a gamma-ray excess in the TS map in the eastern shell of the SNR Puppis A, which we modeled as two point sources (as found by the algorithm), labeled PS J0823.3-4244 (TS$=40.9$) and PS J0823.7-4255 (TS$=82.4$). This indicates that the high-energy emission from the SNR is more complex than given by the simple geometrical model in the 4FGL catalog.

\begin{figure}
 \includegraphics[width=\linewidth]{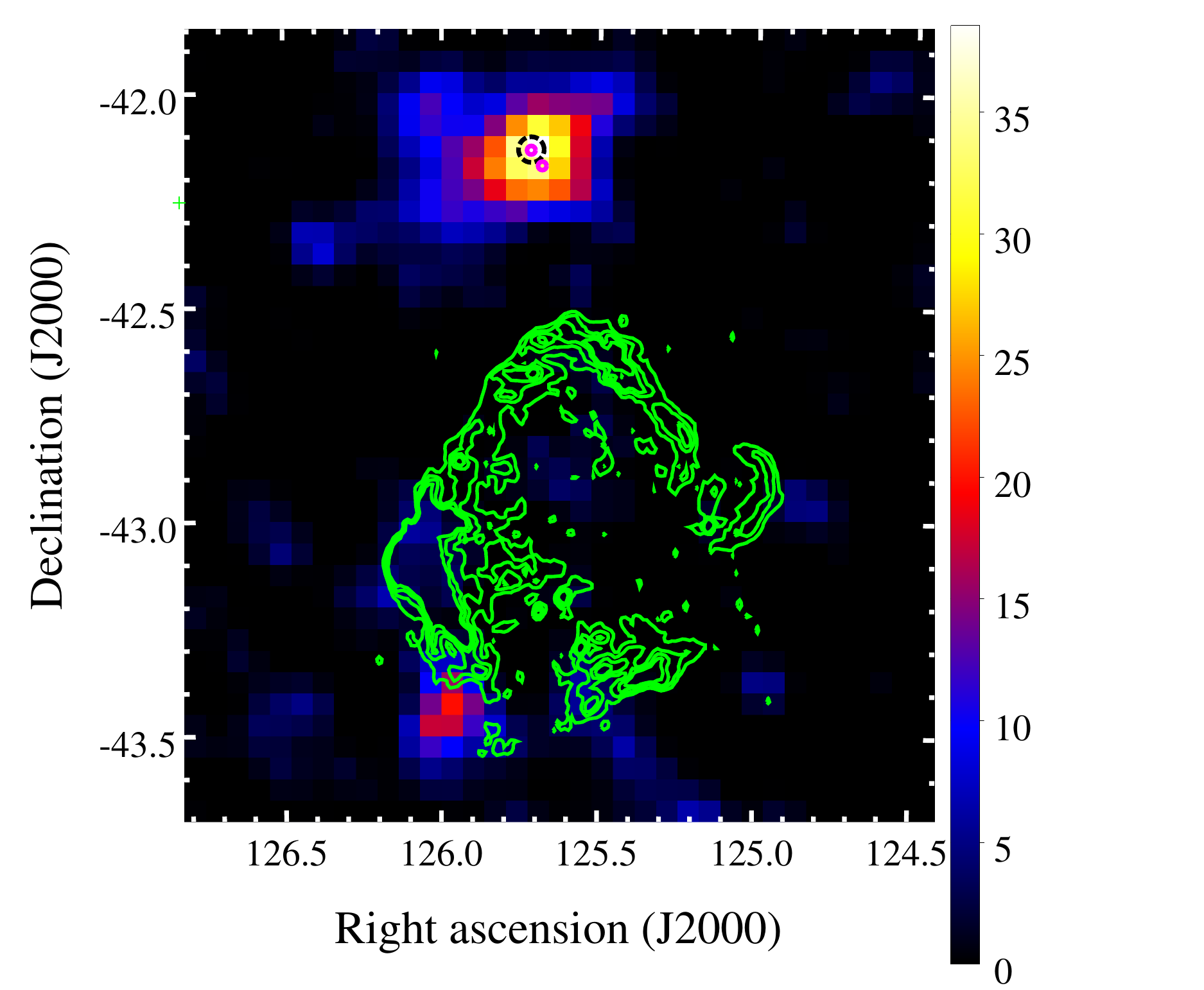}
\caption{TS map obtained with LAT events above an energy of 5 GeV. The gamma-ray emission from the background model is subtracted except for that of \source, which is seen in the image. The dashed black circle represents the 68\% confidence level positional uncertainty region of the GeV source. The magenta circles represent the locations of the star forming regions G259.7592--02.8378 (seen within the error circle of the GeV source) and G259.7743--02.8799. The green contours show the shell of Puppis A, obtained from the Sydney University Molonglo Sky Survey at 843 MHz \citep{2003MNRAS.342.1117M}. \label{fig1:tsmap}}
\end{figure}

We ran the {\tt extension} algorithm on \source using a radial disk as a morphological model and found TS$_{ext} = 6.6$, where TS$_{ext}=2\ln (\mathcal{L}_{ext}/\mathcal{L}_{ps})$, and $\mathcal{L}_{ext}$ and $\mathcal{L}_{ps}$ are the likelihoods resulting from fitting the extended source and a point source, respectively. Detailed simulations show that a reasonable threshold to claim a LAT source as extended is TS$_{ext} > 16$ \citep{Lande_2012}. We concluded that \source is a point source for the LAT and calculated a 95\%-confidence level extension upper limit of $0.19\degr$. A TS map showing \source, together with the locations of the SNR Puppis A and the star formation regions mentioned earlier can be seen in Fig. \ref{fig1:tsmap}.

We repeated the previous analysis using all events with energies above 1 GeV and we found the same point sources near the centre of the ROI as before. We also found that the source corresponding to \source is not significantly extended. The fit above 1 GeV yielded TS$=51.6$ for \source using a simple power-law as the spectral function, which is given by $$\frac{dN}{dE} = N_0 \left( \frac{E}{E_0} \right)^{-\gamma},$$ where $N_0$ is the normalization, $\gamma$ is the spectral index and $E_0$ is a constant scale factor. We then performed a new fit changing the spectral shape of the source to a log-parabola\footnote{See https://fermi.gsfc.nasa.gov/ssc/data/analysis/\\scitools/source\_models.html for the definition.} in order to probe for spectral curvature, and this model yielded TS $=52.6$. Since the two models are nested an increase of the TS value of 1 using one additional degree of freedom indicates that the measured spectrum does not significantly deviate from a simple power-law ($1\sigma$ improvement). The resulting spectral index and integrated photon flux (in the energy range $1-300$ GeV) for the source, with their $1\sigma$ statistical uncertainties, are $\gamma = 2.02\pm0.01$ and $\int \frac{dN}{dE}\,dE = (4.4\pm0.3)\times 10^{-10}$ cm$^{-2}$ s$^{-1}$, respectively, which are consistent with the values reported in the 4FGL catalog ($2.05\pm 0.19$ and $(3.7\pm1.1)\times 10^{-10}$ cm$^{-2}$ s$^{-1}$, respectively).

In order to obtain spectral flux points, we divided the data in the $1-300$ GeV energy range into six logarithmically-spaced intervals and fitted the normalization of the source of interest as well as those of the background sources in each interval. If the TS of the source was below 4 in a given interval we estimated a 95\%-confidence level upper limit on the flux using the likelihood profile. The spectral energy distribution (SED) points plotted together with the overall fit in the $1-300$ GeV energy range are shown below.

\section{Swift-XRT data}
\label{XRT}

To investigate HH219 in more detail as a potential source of the gamma-ray emission from 4FGL J0822.8--4207, we submitted a Target of Opportunity request to the Niel Gehrels \textit{Swift} Observatory (henceforth \textit{Swift}). Observations with \textit{Swift} are a powerful tool for localizing high-energy emission from \textit{Fermi}-LAT sources by using the \textit{Swift} X-ray Telescope \cite[\textit{Swift}-XRT,][]{Burrows_2005} to observe $0.3-10.0$ keV X-rays \citep[for example, the ongoing \textit{Swift} follow-up program described in ][]{2021AJ....161..154K}.  As many sources of gamma rays observed with \emph{Fermi} also produce X-rays via synchrotron emission, if HH219 is producing gamma-ray emission by leptonic processes, it may produce characteristic X-rays as well. Observations of HH219 lasted approximately 2.5 ks.

\begin{figure}
 \includegraphics[width=\linewidth]{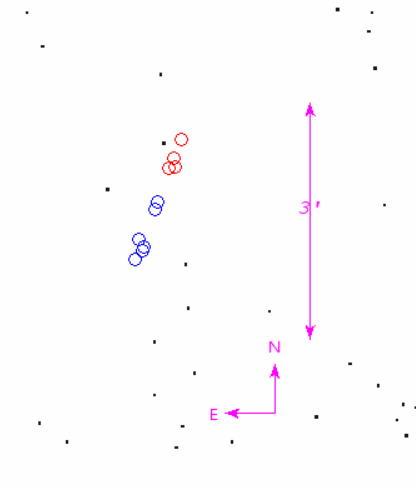}
\caption{\textit{Swift}-XRT event map near HH219. Individual dots are X-ray event detections during the 2.5 ks exposure. Blue and red circles are positions of the blue (approaching, corresponding to HH219) and red knots (receding, unnamed) noted in Table 3 of \protect\cite{2004A&A...422..141C}.\label{fig:xrtmap}}
\end{figure}

Downloading and analyzing the observations using HEASOFT and XSELECT, we produced the event maps shown in Figure \ref{fig:xrtmap}. In our observations, there is no detected X-ray source at the position of HH219. This may be an indication that there is no X-ray emission emerging from HH219 in the $0.3-10.0$ keV range observed by \textit{Swift}-XRT, which has implications about the type of particles being accelerated if HH219 is posited as the source of the \textit{Fermi}-LAT gamma-ray emission. Alternatively, as HH219 is located near a prominent star-forming region, the unusually dense interstellar material in the region may contribute to  large hydrogen column density, absorbing X-rays while admitting gamma rays. We estimated the hydrogen column density in the direction of the source, using the High Energy Astrophysics Science Archive Research Center nH tool \citep{1990ARA&A..28..215D,2005A&A...440..775K}, to be $8.8\times 10^{21}$ cm$^{-2}$. Investigating optical extinction in the same region with Sloan Digital Sky Survey data \citep{2011ApJ...737..103S} we obtained $A_V=6.258$ mag which corresponds to an approximate hydrogen column density of $1.25\times 10^{22}$ cm$^{-2}$. Using a conversion given by \cite{2000ApJ...542..914W} to estimate the photoionization cross section ($\sigma$) in the interstellar medium at various energies, we found that at 0.5 keV $\sigma \sim 10^{-21}$ cm$^2$ per hydrogen atom. Multiplying this cross section with the hydrogen column density gives a rough estimate of an optical depth of the order of 10. This shows that the interstellar column density is more than sufficient to obscure any soft X-rays coming from the direction of HH219. This is a conservative estimate as any additional intrinsic absorption attributed to the source could prevent X-rays with higher energies from escaping. Regardless of explanation, there are not clear-cut X-rays coming from HH219 that could help discern emission mechanisms in the region.

\section{Origin of the gamma rays}\label{model}
\subsection{CR illumination from Puppis A}
Recent studies of the GeV emission from the SNR Puppis A favor a hadronic origin for the gamma rays \citep{2017ApJ...843...90X}, which implies this SNR is a potential source of CRs. We adopt the model by \cite{2011MNRAS.410.1577O} to account for the GeV emission of \source as a result of interactions between high-energy CRs that escaped from Puppis A in the past and interact with dense gas known to exist at the location of this gamma-ray source. Spherically symmetric particle escape is assumed. The absence of X-ray emission from \source is consistent with a hadronic scenario for the gamma rays. We model the SNR as a sphere with a subtended radius of $0.45\degr$ whose centre is at an angular distance of $0.87\degr$ away from \source. The distribution of CRs with momentum $p$ at a distance $r$ from the SNR centre and at a given time $t$ from the explosion, $f(t,r,p)$, is given by the diffusion equation
\begin{equation}
\frac{\partial f(t,r,p)}{\partial t} - D_{\mbox{\tiny ISM}}(p) \Delta f(t,r,p) = q_s(t,r,p),
\label{eq1}
\end{equation}
where $D_{\mbox{\tiny ISM}}(p)$ is the diffusion coefficient in the interstellar medium and $q_s$ is the CR source term. The appropriate source term for particles escaping from the SNR surface at a distance $R_{\mbox{\tiny esc}}(p)$ from the centre at the instant $t_{\mbox{\tiny esc}}(p)$ \citep{2010ApJ...721L..43O,2011MNRAS.410.1577O} is $$q_s = \frac{N_{\mbox{\tiny esc}}(p)}{4 \pi r^2}\,\delta[r - R_{\mbox{\tiny esc}}(p)]\,\delta [t-t_{\mbox{\tiny esc}}(p)],$$ where $N_{\mbox{\tiny esc}}(p)$ is the momentum distribution of the accelerated CRs that escape from the SNR. The solution to equation (\ref{eq1}) can be written as

\begin{equation}
f(t,r,p) = \frac{\mbox{e}^{-\left(\frac{r-R_{\mbox{\tiny esc}}(p)}{R_d(t,p)} \right)^2} - \mbox{e}^{-\left(\frac{r+R_{\mbox{\tiny esc}}(p)}{R_d(t,p)} \right)^2}}{4\pi^{3/2} R_d(t,p) R_{\mbox{\tiny esc}}(p)\, r }\,N_{\mbox{\tiny esc}}(p), \label{eq2}
\end{equation}

with $$R_d = \sqrt{4 D_{\mbox{\tiny ISM}}(p) [t-t_{\mbox{\tiny esc}}(p)] }.$$ Given that the gamma-ray source \source is a point source for the LAT we can obtain the momentum distribution of CRs at the location of this cloud after multiplying the result in equation (\ref{eq2}) by the volume of the emission region. As mentioned earlier, we derived an upper limit on the GeV region size of $0.19\degr$, which we use as the corresponding radius of the source.

As is common in the literature, we parameterize the spectrum of CRs that escape into the interstellar space as $$N_{\mbox{\tiny esc}}(p) = k\, p^{-s} \mbox{e}^{-p/p_{\mbox{\tiny max}}},$$ with $k$ the normalization constant, $s$ the spectral index of the distribution and $p_{\mbox{\tiny max}}$ the maximum momentum of the particles reached at some point in the history of the SNR. The time when a particle escapes from the SNR depends on its diffusion length and, in turn, on the diffusion coefficient near the shock front, which is dependent on the level of turbulence generated by the CRs themselves. Since considering these factors would make the problem very difficult to solve, a phenomenological approach is usually adopted \citep[see, e.g.,][]{2009MNRAS.396.1629G,2011MNRAS.410.1577O,2014ApJ...786..145H}. CRs with the maximum possible energy $p_{\mbox{\tiny max}}$ are assumed to escape when the SNR enters the Sedov-Taylor phase of evolution and has a radius $R_{\mbox{\tiny Sedov}}$. From then on the momenta of escaped particles follows a power-law dependence on the radius as $$p = p_{\mbox{\tiny max}} \left( \frac{R_{\mbox{\tiny esc}} }{R_{\mbox{\tiny Sedov}} } \right)^{-\alpha}.$$ Although some authors usually take $p_{\mbox{\tiny max}} = 10^{15.5}$ eV$/c$, corresponding to the break in the locally observed CR energy spectrum known as ``the knee'', we note that no SNR has been proven to accelerate particles to these energies and, somewhat surprisingly, even young SNRs show TeV spectra that instead cutoff at very low energies compared to those expected from CRs reaching the knee. A famous example of an SNR that could be near the Sedov-Taylor phase is Cas A, which shows a cutoff in its TeV spectrum around $2.3$ TeV \citep{2020ApJ...894...51A}. We therefore take $p_{\mbox{\tiny max}}$ in the range 40--100 TeV $c^{-1}$, as required by our fits to the data. The value of the parameter $\alpha$ has been taken as 6.5 in the literature \citep{2010ApJ...721L..43O} so that $p = 1$ GeV$/c$ at the end of the Sedov-Taylor phase, but we also treat it as a free parameter to be determined by the best fit to the GeV data. Using the Sedov solution it is also found that \citep{2011MNRAS.410.1577O} $$t_{\mbox{\tiny esc}} = t_{\mbox{\tiny Sedov}} \left( \frac{p}{p_{\mbox{\tiny max}}} \right)^{-5/2\alpha}.$$ 

When fitting the data, this equation allows us to estimate the current age of Puppis A given the momentum of the CRs currently escaping from the SNR, which we also take as a free parameter. In order to be able to explain the gamma-ray flux from \source around 1 GeV, the energies of the cosmic rays that are currently escaping from the SNR should be in the 5--10 GeV range. In Section \ref{snrspectrum} below we discuss the impact this may have on the gamma-ray spectrum from the SNR itself. Other free parameters in our treatment are the spectral index $s$ (varied in the range 2--2.2) and normalization $k$ of the injected particles, the distance $r$ between the centre of the SNR and the GeV source, and the parameters $\delta$ and $D_{27}$ describing the diffusion coefficient in the interstellar medium, which is given by $$D_{\mbox{\tiny ISM}}(p) = 10^{27}D_{27} \left( \frac{p}{10 \,\mbox{GeV}c^{-1}}  \right)^\delta\,\,\,\,\mbox{cm$^2$ s$^{-1}$}.$$ Note that given the observed SNR radius (10 pc for a distance to the source of 1.3 kpc) $\alpha$, $p_{\mbox{\tiny max}}$ and the momentum of the particles currently escaping from the SNR, the Sedov radius can also be calculated, and the parameters were varied keeping its value in the range 1.9--3 pc. Since the model has a large number of free parameters, the Sedov time was fixed at 180 yr in all calculations. Since the observed radius of the SNR, the radius of the gamma-ray emission region, the distance between the SNR and the cloud and the flux of gamma rays measured on Earth depend on the distance to the observer, $d$, we repeated the calculations considering several values of $d$ within the uncertainty interval reported recently for Puppis A, which allows us to derive some constraints on the distance between the SNR and \source under this model for the origin of the gamma rays.

Finally, we used the parameterization of the photon production cross section for proton-proton interactions presented by \cite{2014PhRvD..90l3014K} and the {\tt naima} package \citep{naima} for the calculation of the photon fluxes and total energy in the particles. The proton number density of the dense clump at the location of the GeV source was estimated by \cite{2005A&A...432..921F} as $n=2\times 10^5$ cm$^{-3}$. This large density is likely seen only in a very small region of space. We first calculated the energy content in the cosmic rays at the location of \source as a function of $n$, which is seen below. Secondly, by imposing a reasonable constraint on the total energy injected to all the cosmic rays by the SNR, we estimate the necessary values of the density in the target material at the location of \source.

\subsubsection{Results}
\begin{figure}
 \includegraphics[width=10cm]{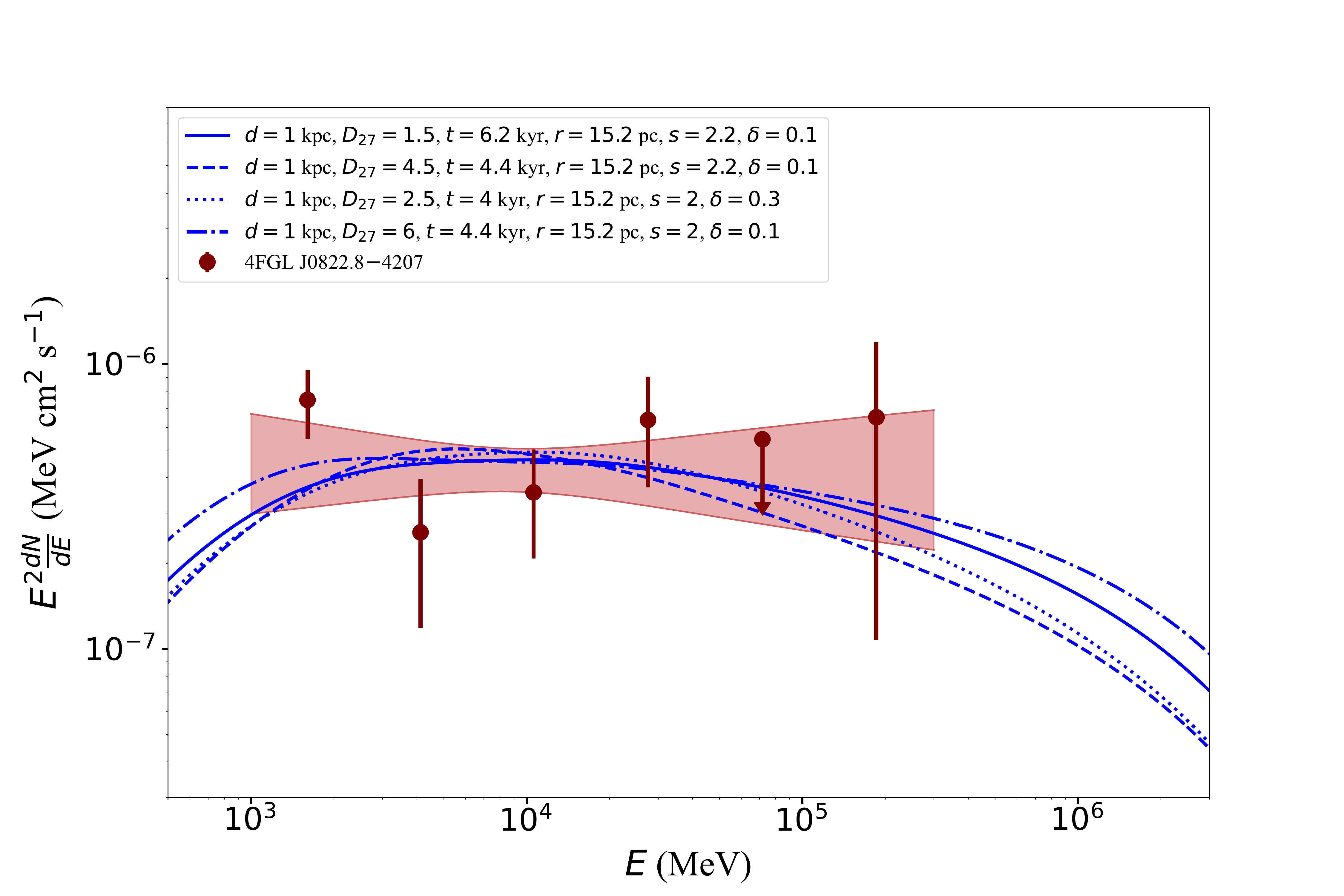}
 \includegraphics[width=10cm]{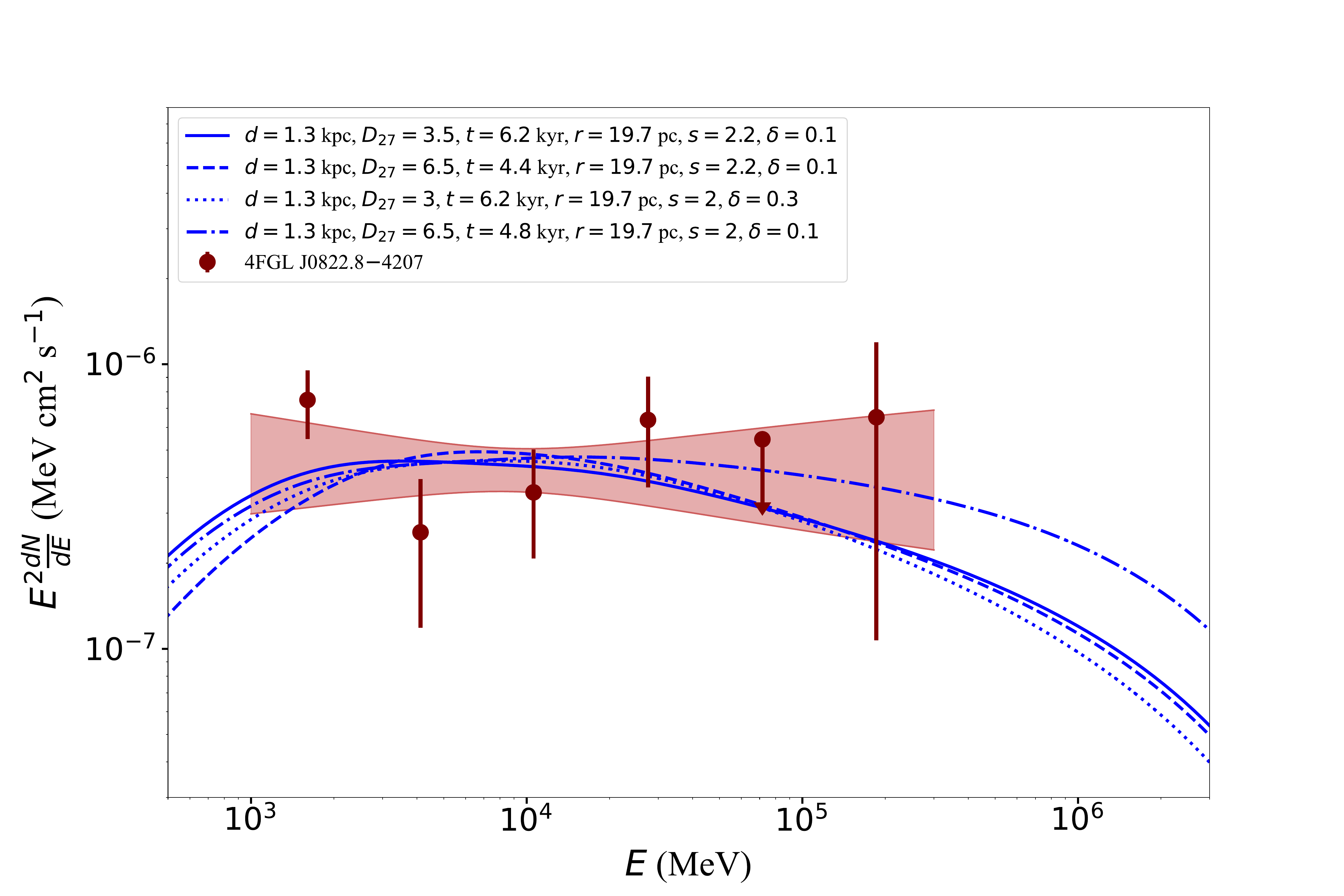}
 \includegraphics[width=10cm]{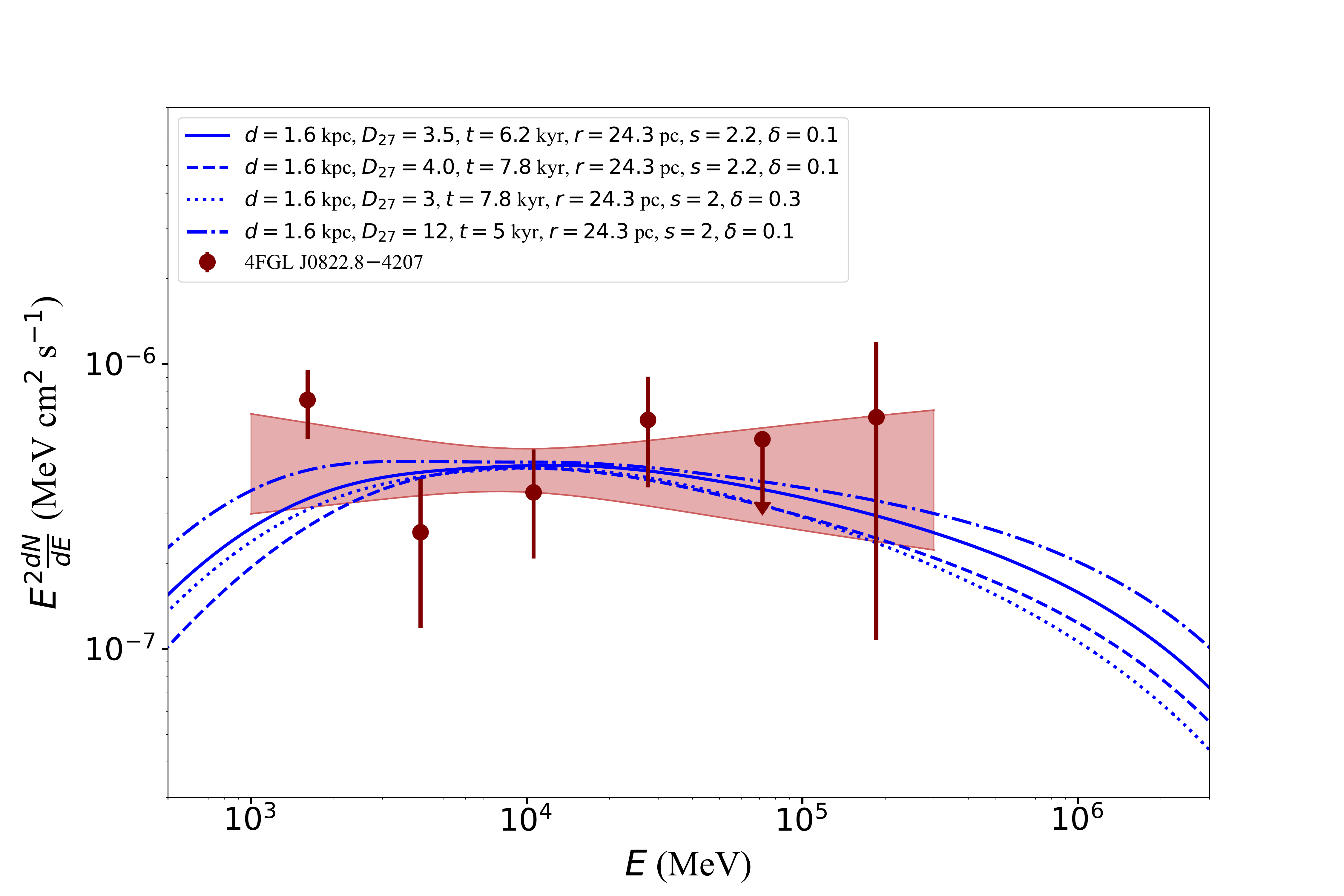}
\caption{SED of \source obtained in this work. The flux points estimated in Section \ref{LAT} are shown as well as the propagated $1\sigma$ uncertainty band (shaded region) of the spectral fit in the 1--500 GeV energy range. The predicted fluxes from CR illumination from Puppis A are shown by the solid, dashed, dotted and dash-dotted lines for different model parameters. The parameter values are shown for three possible distances to the source, from top to bottom, $d=$1, 1.3 and 1.6 kpc. \label{fig2:CRmodel}}
\end{figure}

The results for the CR-illumination scenario are seen in Fig. \ref{fig2:CRmodel} where the high-energy SED of \source obtained in section \ref{LAT} is compared to the model predictions for different parameter sets. These parameters are shown in table \ref{table1}. The results are shown for three possible distances ($d$) to the system, 1, 1.3 and 1.6 kpc and $s=2.2$ and 2.0. The model explains the observed fluxes reasonably well in all cases. The values of $r$ shown correspond to the lowest possible distance between the cloud and the SNR, obtained directly from their angular separation in the sky. The effect of increasing $r$ is discussed below.

\begin{table*}
\centering
\caption{Sets of possible parameter values found in the CR-illumination scenario for the origin of \source.}
\label{table1}
\begin{tabular}{lccccccc}
\hline\hline
$d$ (kpc) & $t$ (kyr)$^a$ & $\alpha$ & $D_{27}$ & $\delta$ & $r$ (pc) & $s$ & $W_p \cdot \left( \frac{n}{1 \,\mbox{\tiny cm$^{-3}$}} \right)$ (erg)$^b$\\
\hline
1         &     4.4       &  6.5     &   4.5    &    0.1   & 15.2    & 2.2  & $2.2\times 10^{48}$\\
          &     6.2       &  6.5     &   1.5    &    0.1   & 15.2    & 2.2  & $2.3\times 10^{48}$\\
          &     4         &  7.4     &   2.5    &    0.3   & 15.2    & 2    & $2.2\times 10^{48}$\\
          &     4.4       &  7.2     &   6      &    0.1   & 15.2    & 2    & $2.7\times 10^{48}$\\
\hline
1.3       &     4.4       &  6.5     &   6.5    &    0.1   & 19.7    & 2.2 & $3.6\times 10^{48}$ \\
          &     6.2       &  6.5     &   3.5    &    0.1   & 19.7    & 2.2 & $4.0\times 10^{48}$ \\
          &     6.2       &  6.5     &   3      &    0.3   & 19.7    & 2   & $3.6\times 10^{48}$ \\
          &     4.8       &  7     &   6.5      &    0.1   & 19.7    & 2   & $4.4\times 10^{48}$ \\

\hline
1.6       &     7.8       &  5.5     &   4.0    &    0.1   & 24.3    & 2.2 & $4.8\times 10^{48}$\\
          &     6.2       &  6.5     &   3.5    &    0.1   & 24.3    & 2.2 & $5.7\times 10^{48}$\\
          &     7.8       &  6.1     &   3      &    0.3   & 24.3    & 2   & $5\times 10^{48}$\\
          &     5         &  6.9     &   12     &    0.1   & 24.3    & 2   & $6.7\times 10^{48}$\\
\hline
\end{tabular}\\
\textsuperscript{$a$}\footnotesize{Age of Puppis A.}\\
\textsuperscript{$b$}\footnotesize{The product of the total energy in the cosmic rays at the location of \source and the number density of the target gas in cm$^{-3}$.}
\end{table*}

Several trends and properties of the model can be summarized as follows:
\begin{enumerate}
\item Increasing the value of $\alpha$ results in a harder predicted spectrum, as it also lowers the predicted age of the remnant for a fixed momentum of escaping particles.
\item An increase in the diffusion coefficient normalization, $D_{27}$, results as expected in an increase of the fluxes, particularly at the lowest energies.
\item For fixed spectral parameters, lowering the distance to the system, $d$, also makes the spectrum softer (increasing fluxes at the lowest energies).
\item Increasing the distance from the cloud to the SNR, $r$, produces harder predicted spectra as only higher-energy particles are able to reach the cloud.
\item Values of $\delta$ (which determines the energy dependence of the diffusion coefficient and, therefore, the particle and photon spectral index at the source) higher than $\sim 0.3$ result in low fluxes at higher energies, inconsistent with data. If the injected CR spectrum has a spectral index $s=2.2$ and the SNR age is chosen to be $\lesssim 4$ kyr for $d=1$ kpc and $d=1.3$ kpc, values around $\delta=0.2$ are allowed. For $d=1.6$ kpc, a relatively good fit to the data can be obtained for $\delta \sim 0.2$, $s=2.2$ and an age $\sim 6.5$ kyr, but the Sedov-Taylor radius increases above 3 pc, outside the range of values considered here as plausible. For a harder injected spectrum with $s=2$, reasonable predicted fluxes are obtained for $\delta=0.3$.
\end{enumerate}

As seen in Table \ref{table1} the values of $D_{27}$ are smaller than those derived from propagation models in the Galaxy \citep[$\sim10$, e.g.,][]{Ptuskin_2006}. A similar situation was found in the W28 system \citep{2014ApJ...786..145H}. Low values of the diffusion coefficient in the vicinity of an SNR could be caused by magnetic turbulence generated by the escaping CRs \citep[e.g.,][]{2016MNRAS.461.3552N}. As expected, since the gamma-ray spectrum of \source is relatively hard (with a spectral index $\sim 2$), low values of $\delta \sim 0.1-0.2$ are favored in our analysis, while the typical values in the Galaxy derived from CR propagation are $\sim 0.6$ \citep{Ptuskin_2006}. The maximum allowed values of $\delta$ are $\sim0.3$ and they generally require an older remnant and a harder injected spectrum for the CRs with a spectral index of $\sim2$. The obtained values of $\delta$ are similar to those in the vicinity of the W28 SNR found with the same model \citep{2014ApJ...786..145H}, which are lower than the average Galactic values. The normalization of the diffusion coefficient, as well as $\delta$, could be modified by the amplification of Alfv\'en waves by cosmic rays \citep{2011MNRAS.415.3434F,2013ApJ...768...73M}. A modification is also expected even if the waves suffer from different types of damping \citep{2021arXiv210604948R}.

The predicted fluxes of gamma rays in the TeV range increase, as expected, for harder injected spectra and lower values of $\delta$. Very-high energy observations could be used to probe these scenarios. A higher distance, $d$, to the system generally requires an older SNR, or a combination of harder particle spectra ($s=2$, $\delta=0.1$) and higher diffusion coefficient ($D_{27} \sim 12$) for a younger remnant ($t=5$ kyr). Some of these trends can be seen in Fig. \ref{fig2:CRmodel}.

Using the normalization of the spectrum of cosmic rays that is required to explain the emission at the location of \source, we calculated the total energy that is transferred to the relativistic particles that are injected by the SNR, $W_T$, above the particle energies that are constraint by the data ($5-10$ GeV). In terms of the number density of the target gas at the location of \source, the values vary in the range $$W_T = (0.9-3.2)\times 10^{51} \,\mbox{erg}\, \left( \frac{1\,\mbox{cm}^{-3}}{n} \right)$$ for the scenarios presented in Table \ref{table1}. Since the total energy injected into cosmic rays by an SNR is expected to be of the order of $10^{50}$ erg, integrated above particle energies of a few GeV, we constrain the target density of the gas at the location of the source to be $n>9-32$ cm$^{-3}$, depending on the model parameters shown in Table \ref{table1}. The required densities could be larger if the total energy injected into cosmic rays is less than 10\% of the typical kinetic energy in the shock of an SNR or for a supernova that is less energetic. This is not an issue given the existence of very dense clumps of gas at the location of \source.

These results were obtained using the minimum allowed value for the distance between the SNR and the cloud, $r$, determined by their angular separation in the sky and the distance to the system ($d$). We also studied the effect of increasing $r$ and found that the predicted fluxes from this model can still account for the gamma-ray observations if the normalization of the diffusion coefficient is increased and the age of the remnant is $\gtrsim 6$ kyr (for $s=2.2$). Fixing the diffusion coefficient to $D_{27}=10$, derived from propagation models in the Galaxy, a maximum age for Puppis A of 7 kyr, $\delta=0.1$ and $s=2.2$, the maximum values for $r$ become: 35 pc for $d=1$ kpc and 40 pc for $d=1.3$ kpc and $d=1.6$ kpc. Note that the SNR radius is also dependent on the distance to the system and this is adjusted accordingly. If instead a harder spectrum is used with $s=2$, much higher fluxes are predicted at TeV energies but a larger value of $r$ is not necessarily allowed because lower energy CRs start to require $D_{27}>10$ to reach the target gas and reproduce the SED at the lowest energies. This is also a plausible scenario, of course, where higher values of $r$ are possible for a higher diffusion coefficient and $s=2$.

\subsubsection{On the GeV spectrum of Puppis A}\label{snrspectrum}
The results presented so far assume the energies of protons currently escaping from the SNR to be in the range 5--10 GeV. This is necessary in order to explain the flux from \source at the lowest energies, around 1 GeV. If particles can no longer be confined within an SNR above certain energy, this is expected to produce a break in its gamma-ray spectrum \citep[e.g.,][]{2010Sci...327.1103A,2011NatCo...2..194M,2019MNRAS.490.4317C}. A recent analysis of the GeV spectrum of Puppis A by \cite{2020PASJ...72...72S} found no sign of a spectral break, while a previous work by \cite{2017ApJ...843...90X} found a spectral break around $\sim 8$ GeV with a relatively steep spectrum with an index of $\sim2.5$ above this break energy, which would also explain the lack of TeV emission from Puppis A \citep{2015A&A...575A..81H}.

The escape of cosmic rays with energies in the 5--10 GeV range would translate into a break in the photon spectrum of the SNR around 2 GeV \citep{2010Sci...327.1103A,2011NatCo...2..194M}, which has not been found. Future observations with more statistics could help confirm or discard the model presented here. If, on the other hand, the photon spectral break around 8 GeV is confirmed for Puppis A and if it is caused by particle escape, the associated particle escape energies would be of the order of tens of GeV. We found that when fixing the energy of currently escaping cosmic rays above $\sim 20$ GeV it is not possible to explain the emission from \source at the lowest energies with reasonable model parameters. However, the model can explain the emission from \source above photon energies of $\sim4$ GeV assuming that the currently escaping particles have energies above several tens of GeV. This would be consistent with a scenario where the observed break in the SNR spectrum is produced by particle escape. In this case, the flux from \source at lower energies could result from a different physical process, perhaps associated to star formation, as discussed in the next section.

Models of particle escape and confinement near supernova remnants, incorporating the properties of the particles and of the waves they generate through the cosmic ray streaming instability, predict that for an SNR age of $\sim5$ kyr the energies of escaping particles should be in the $\sim100-1000$ GeV range \citep{2016MNRAS.461.3552N}. Therefore explaining the spectrum of \source by illumination of cosmic rays from Puppis A would require a much more rapid escape of particles than predicted by these models, and perhaps less efficient turbulence generation. Alternatively, as mentioned earlier, a scenario where particles currently escaping have energies greater than a few tens of GeV would only explain the higher energy gamma-ray fluxes seen from \source.

\subsection{A protostellar jet}
As mentioned earlier, a Herbig-Haro object (HH219) in the complex IRAS 08211--4158 is known to exist at the location of the GeV source \source \citep{1991PASP..103...79G}. Detailed observations of the jet, which is associated with the youngest and most massive object in the cluster, a very young B2 spectral type star of 642$L_\odot$ and $5-6$ $M_\odot$ \citep{2004A&A...422..141C}, were carried out by \cite{2002ApJ...564..839L} and \cite{2004A&A...422..141C}. They identified $24$ knots moving along the jet and bow shock structures that are visible around some of them, as well as inner knots indicating that the activity of the outflow is still ongoing. Using their average values for the knot sizes ($1.4\times 10^{-2}$ pc), densities ($5\times10^3$ cm$^{-3}$), velocities (300 km s$^{-1}$) and dynamical ages (300 yr) we estimated an order of magnitude of the kinetic luminosity of the jet, $\frac{1}{2}\dot{M} v^2$, of $10^{35}$ erg s$^{-1}$, similar to that of IRAS 16547--4247 \citep{2010A&A...511A...8B}. The parameters used for this estimate were obtained on the assumption that the system is located at 400 pc. Based on our data analysis, we estimated the gamma-ray luminosity of \source for energies above 1 GeV of $7.4\times10^{31}$ erg s$^{-1}$, assuming also a distance of 400 pc to the source. This value is consistent with predicted estimates \citep[e.g.,][]{2012AIPC.1505..281A}. The gamma-ray luminosity is thus only a small fraction of the available kinetic energy and this conclusion would not be affected if instead an actual source distance estimate of 1.7 kpc was used \citep{2005A&A...432..921F}. For an increase by a factor of 4 in the distance to the system, the gamma-ray luminosity would be $\sim10^{33}$ erg. This makes the production of non-thermal emission in the HH219 system energetically feasible. The observed gamma-ray luminosity is consistent with theoretical predictions \citep{2010A&A...511A...8B} for the Bremsstrahlung scenario from electrons and from proton-proton collisions in the hadronic scenario for the GeV emission. Our observations show an absence of X-ray emission from HH219 which would be compatible with a hadronic scenario for the origin of \source. Alternatively, any X-ray emission from accelerated electrons in HH219 could be absorbed in the dense gas associated to the star-forming region or in its path to Earth. Future observations searching for non-thermal radio emission would be useful in the future to confirm or reject any leptonic contribution.

Previous estimates of the maximum energy that could possibly be reached by particles in a similar protostellar jet, associated to IRAS 16547--4247, are of the order of 3 TeV for electrons and 50 TeV for protons \citep{2007A&A...476.1289A}, under the assumption that the maximum energy $E_m$ gained in the acceleration process is limited by the confinement in the acceleration region of size $r\sim 10^{16}$ cm, $$\frac{E_m}{\eta e Bc} = \frac{r^2}{D},$$ where $B\sim 1$ mG is the magnetic field, $\eta \sim (v_s/c)^2$ with $v_s \sim 10^3$ km s$^{-1}$ the shock speed and $D$ the (Bohm-limit) diffusion coefficient. These values for the parameters are typical in the environments of massive young stellar objects \citep[see, e.g.,][]{2010A&A...511A...8B}. Given the high number densities in the cloud core ($\sim 10^5$ cm$^{-3}$), this assumption is supported by estimates of the cooling time of the particles which is lower than the diffusion time in the emitting region \citep{2007A&A...476.1289A}. For electrons, the losses are dominated by bremsstrahlung emission, while for relativistic protons they are caused by inelastic collisions with ambient protons in the dense environment of the cloud.

Our derived gamma-ray spectrum of the source \source is consistent with a simple power-law with a spectral index $\sim 2$ which would imply a similar spectral index in the particle distribution for the case of leptonic emission through non-thermal Bremsstrahlung as well as for hadronic emission from protons. This value of the spectral index matches the predictions of standard DSA theory \citep[e.g.,][]{1983RPPh...46..973D}. Given that there is no indication of a cutoff in the photon spectrum at sub-TeV energies, observations of the source by imaging atmospheric Cherenkov telescopes are encouraged to probe the maximum energies currently achieved by the particles. The escape of the highest energy protons could produce gamma-ray emission in the cloud that could also be probed by more detailed observations.

As pointed out before, observations searching for non-thermal radio emission from electrons would help to investigate the nature of the high-energy particles at the source. Since the emission from HH objects can be variable in a timescale of years, future gamma-ray observations with more statistics could reveal a variability that would confirm HH219 as the origin of \source. Studies of GeV and TeV emission from HH objects could help understand the conditions for the formation of massive stars, the process of particle acceleration and the role of these objects as sources of high-energy particles in the Galaxy. If confirmed as a source of relativistic particles, HH219, located in a region of relatively low gamma-ray diffuse Galactic emission compared to regions closer to the Galactic centre, and having a hard GeV spectrum, offers an exciting opportunity in high-energy astronomy.

Finally we point out that it is of course possible that the emission from \source is the result of a combination of the two scenarios explored in this work. For example, cosmic ray illumination from Puppis A could be responsible for only the highest energy fluxes, particularly if the photon break in the SNR spectrum (if confirmed around 8 GeV) is produced by currently escaping cosmic rays with energies above several tens of GeV.

%gamma-ray luminosity (2.4e-6*1e6/(6.2415e11))*4*3.14*(400.0*3.086e18)**2
%%kinetic luminosity = ((2*m.pi/3)*24*((300000.0)**2)*(1.67e-27*5000.0)*((0.7e-2)*(3.086e18))**3/(300.0*31e6))/1e-7

\section{Summary}\label{summary}
Clumps of very dense gas are seen at the location of the unidentified GeV source \source having a systemic velocity that is consistent with recent corresponding estimates for Puppis A. We confirmed the point-like nature of \source, as seen by the LAT, and its relatively hard GeV spectrum, which is described by a simple power-law with a spectral index of $\sim 2$. We have shown that the gamma-ray emission from \source can be explained by cosmic ray illumination from the SNR Puppis A, a process in which high-energy CRs accelerated in the shell of the SNR have escaped and diffused in space to reach the clouds of gas where hadronic interactions produce the observed gamma rays. The lack of X-ray emission shown by our observations is consistent with a hadronic nature for \source. This scenario requires for the gas to be located at a distance of $\lesssim 40$ pc from the centre of the SNR, unless Puppis A is older than previously estimated or the diffusion coefficient is larger than the average Galactic value. Values of $\delta \lesssim 0.3$, defining the energy-dependence of the diffusion coeffcicient in the interstellar medium, are preferred by the data. Additionally, the required energies of particles that are currently escaping from the SNR should be in the range 5--10 GeV if cosmic ray illumination is to account for the flux of \source at the lowest photon energies, around $\sim1$ GeV. If this is true, a photon spectral break is expected in the spectrum of the SNR around 2 GeV according to our model. A possible break in the photon spectrum of the SNR has been reported by \cite{2017ApJ...843...90X} around 8 GeV. If this break is real and produced by cosmic rays that are currently escaping from the SNR (with energies of tens of GeV), then the cosmic-ray illumination scenario could only account for the flux from \source at photon energies above $\sim4$ GeV. We note that the validity of the cosmic ray illumination hypothesis as an origin of any gamma rays from \source depends on the assumption that Puppis A is located at a distance of several tens of pc from the gas. An accurate determination of the distance to these objects should be obtained with additional observations.

A very interesting scenario for the origin of \source is the acceleration of particles in the termination shock of the Herbig-Haro object HH219, which is seen within the 68\% confidence level positional uncertainty region of the GeV source. The kinetic energy of the jet is greater by several orders of magnitude than the energy in the gamma rays and the GeV spectrum is consistent with standard diffusive shock acceleration predictions. The maximum energy predicted for the particles accelerated in these objects is consistent with our GeV data, as well as the observed luminosity. We encourage follow-up observations which could confirm this scenario and reveal HH219 as an excellent laboratory to study particle acceleration.
 
\section*{Acknowledgements}

We thank the anonymous referee for helpful comments, and Y. Ohira for general comments on the cosmic ray escape model. MA and LG have received funding from Universidad de Costa Rica (UCR) and its Escuela de F\'isica under grant number B8267. SK gratefully acknowledges the support of NASA grants 80NSSC17K0752 and 80NSSC18K1730. Thanks to CICIMA-UCR for providing us with computational resources during COVID-19 lockdowns. This research is based on observations made with NASA's Fermi Gamma-Ray Space Telescope, developed in collaboration with the U.S. Department of Energy, along with important contributions from academic institutions and partners in France, Germany, Italy, Japan, Sweden and the U.S.

\section*{Data availability.} The Fermi-LAT data underlying this article are available in the Fermi Science Support Center, at https://fermi.gsfc.nasa.gov/ssc/. X-ray analysis used of data and/or software provided by the High Energy Astrophysics Science Archive Research Center (HEASARC), which is a service of the Astrophysics Science Division at NASA/GSFC. Swift-XRT data used herein is available at the NASA HEASARC archive at https://heasarc.gsfc.nasa.gov/, under observation ID 00014787001. The derived data generated in this research will be shared on request to the corresponding author.

\bibliographystyle{mnras}
\bibliography{references}

\begin{thebibliography}{}
\makeatletter
\relax
\def\mn@urlcharsother{\let\do\@makeother \do\$\do\&\do\#\do\^\do\_\do\%\do\~}
\def\mn@doi{\begingroup\mn@urlcharsother \@ifnextchar [ {\mn@doi@}
  {\mn@doi@[]}}
\def\mn@doi@[#1]#2{\def\@tempa{#1}\ifx\@tempa\@empty \href
  {http://dx.doi.org/#2} {doi:#2}\else \href {http://dx.doi.org/#2} {#1}\fi
  \endgroup}
\def\mn@eprint#1#2{\mn@eprint@#1:#2::\@nil}
\def\mn@eprint@arXiv#1{\href {http://arxiv.org/abs/#1} {{\tt arXiv:#1}}}
\def\mn@eprint@dblp#1{\href {http://dblp.uni-trier.de/rec/bibtex/#1.xml}
  {dblp:#1}}
\def\mn@eprint@#1:#2:#3:#4\@nil{\def\@tempa {#1}\def\@tempb {#2}\def\@tempc
  {#3}\ifx \@tempc \@empty \let \@tempc \@tempb \let \@tempb \@tempa \fi \ifx
  \@tempb \@empty \def\@tempb {arXiv}\fi \@ifundefined
  {mn@eprint@\@tempb}{\@tempb:\@tempc}{\expandafter \expandafter \csname
  mn@eprint@\@tempb\endcsname \expandafter{\@tempc}}}

\bibitem[\protect\citeauthoryear{{Abdo} et~al.,}{{Abdo}
  et~al.}{2010}]{2010Sci...327.1103A}
{Abdo} A.~A.,  et~al., 2010, \mn@doi [Science] {10.1126/science.1182787}, \href
  {https://ui.adsabs.harvard.edu/abs/2010Sci...327.1103A} {327, 1103}

\bibitem[\protect\citeauthoryear{{Abdollahi} et~al.,}{{Abdollahi}
  et~al.}{2020}]{2020ApJS..247...33A}
{Abdollahi} S.,  et~al., 2020, \mn@doi [\apjs] {10.3847/1538-4365/ab6bcb},
  \href {https://ui.adsabs.harvard.edu/abs/2020ApJS..247...33A} {247, 33}

\bibitem[\protect\citeauthoryear{{Abeysekara} et~al.,}{{Abeysekara}
  et~al.}{2020}]{2020ApJ...894...51A}
{Abeysekara} A.~U.,  et~al., 2020, \mn@doi [\apj] {10.3847/1538-4357/ab8310},
  \href {https://ui.adsabs.harvard.edu/abs/2020ApJ...894...51A} {894, 51}

\bibitem[\protect\citeauthoryear{{Acero}, {Lemoine-Goumard}, {Renaud},
  {Ballet}, {Hewitt}, {Rousseau}  \& {Tanaka}}{{Acero}
  et~al.}{2015}]{2015A&A...580A..74A}
{Acero} F.,  {Lemoine-Goumard} M.,  {Renaud} M.,  {Ballet} J.,  {Hewitt} J.~W.,
   {Rousseau} R.,   {Tanaka} T.,  2015, \mn@doi [\aap]
  {10.1051/0004-6361/201525932}, \href
  {https://ui.adsabs.harvard.edu/abs/2015A&A...580A..74A} {580, A74}

\bibitem[\protect\citeauthoryear{{Acero} et~al.,}{{Acero}
  et~al.}{2016}]{2016ApJS..224....8A}
{Acero} F.,  et~al., 2016, \mn@doi [\apjs] {10.3847/0067-0049/224/1/8}, \href
  {https://ui.adsabs.harvard.edu/abs/2016ApJS..224....8A} {224, 8}

\bibitem[\protect\citeauthoryear{{Ackermann} et~al.,}{{Ackermann}
  et~al.}{2013}]{2013Sci...339..807A}
{Ackermann} M.,  et~al., 2013, \mn@doi [Science] {10.1126/science.1231160},
  \href {https://ui.adsabs.harvard.edu/abs/2013Sci...339..807A} {339, 807}

\bibitem[\protect\citeauthoryear{{Aharonian} \& {Atoyan}}{{Aharonian} \&
  {Atoyan}}{1996}]{1996A&A...309..917A}
{Aharonian} F.~A.,  {Atoyan} A.~M.,  1996, \aap, \href
  {https://ui.adsabs.harvard.edu/abs/1996A&A...309..917A} {309, 917}

\bibitem[\protect\citeauthoryear{{Aharonian} et~al.,}{{Aharonian}
  et~al.}{2008}]{2008A&A...481..401A}
{Aharonian} F.,  et~al., 2008, \mn@doi [\aap] {10.1051/0004-6361:20077765},
  \href {https://ui.adsabs.harvard.edu/abs/2008A&A...481..401A} {481, 401}

\bibitem[\protect\citeauthoryear{{Aharonian}, {Yang}  \& {de O{\~n}a
  Wilhelmi}}{{Aharonian} et~al.}{2019}]{2019NatAs...3..561A}
{Aharonian} F.,  {Yang} R.,   {de O{\~n}a Wilhelmi} E.,  2019, \mn@doi [Nature
  Astronomy] {10.1038/s41550-019-0724-0}, \href
  {https://ui.adsabs.harvard.edu/abs/2019NatAs...3..561A} {3, 561}

\bibitem[\protect\citeauthoryear{{Araudo} \& {Rodr{\'\i}guez}}{{Araudo} \&
  {Rodr{\'\i}guez}}{2012}]{2012AIPC.1505..281A}
{Araudo} A.~T.,  {Rodr{\'\i}guez} L.~F.,  2012, in {Aharonian} F.~A.,
  {Hofmann} W.,   {Rieger} F.~M.,  eds,  American Institute of Physics
  Conference Series Vol. 1505, High Energy Gamma-Ray Astronomy: 5th
  International Meeting on High Energy Gamma-Ray Astronomy. Heidelberg,
  Germany. pp 281--284 (\mn@eprint {arXiv} {1210.3078}),
  \mn@doi{10.1063/1.4772252}

\bibitem[\protect\citeauthoryear{{Araudo}, {Romero}, {Bosch-Ramon}  \&
  {Paredes}}{{Araudo} et~al.}{2007}]{2007A&A...476.1289A}
{Araudo} A.~T.,  {Romero} G.~E.,  {Bosch-Ramon} V.,   {Paredes} J.~M.,  2007,
  \mn@doi [\aap] {10.1051/0004-6361:20077636}, \href
  {https://ui.adsabs.harvard.edu/abs/2007A&A...476.1289A} {476, 1289}

\bibitem[\protect\citeauthoryear{{Ballet}, {Burnett}, {Digel}  \&
  {Lott}}{{Ballet} et~al.}{2020}]{2020arXiv200511208B}
{Ballet} J.,  {Burnett} T.~H.,  {Digel} S.~W.,   {Lott} B.,  2020, arXiv
  e-prints, \href {https://ui.adsabs.harvard.edu/abs/2020arXiv200511208B} {p.
  arXiv:2005.11208}

\bibitem[\protect\citeauthoryear{{Becker}, {Prinz}, {Winkler}  \&
  {Petre}}{{Becker} et~al.}{2012}]{2012ApJ...755..141B}
{Becker} W.,  {Prinz} T.,  {Winkler} P.~F.,   {Petre} R.,  2012, \mn@doi [\apj]
  {10.1088/0004-637X/755/2/141}, \href
  {https://ui.adsabs.harvard.edu/abs/2012ApJ...755..141B} {755, 141}

\bibitem[\protect\citeauthoryear{{Blair}, {Sankrit}, {Ghavamian}, {Raymond}  \&
  {Morse}}{{Blair} et~al.}{2003}]{2003AAS...203.3912B}
{Blair} W.~P.,  {Sankrit} R.,  {Ghavamian} P.,  {Raymond} J.~C.,   {Morse}
  J.~A.,  2003, in American Astronomical Society Meeting Abstracts. p. 39.12

\bibitem[\protect\citeauthoryear{{Blasi}}{{Blasi}}{2013}]{2013A&ARv..21...70B}
{Blasi} P.,  2013, \mn@doi [\aapr] {10.1007/s00159-013-0070-7}, \href
  {https://ui.adsabs.harvard.edu/abs/2013A&ARv..21...70B} {21, 70}

\bibitem[\protect\citeauthoryear{{Bosch-Ramon}, {Romero}, {Araudo}  \&
  {Paredes}}{{Bosch-Ramon} et~al.}{2010}]{2010A&A...511A...8B}
{Bosch-Ramon} V.,  {Romero} G.~E.,  {Araudo} A.~T.,   {Paredes} J.~M.,  2010,
  \mn@doi [\aap] {10.1051/0004-6361/200913488}, \href
  {https://ui.adsabs.harvard.edu/abs/2010A&A...511A...8B} {511, A8}

\bibitem[\protect\citeauthoryear{{Burrows} et~al.,}{{Burrows}
  et~al.}{2005}]{Burrows_2005}
{Burrows} D.~N.,  et~al., 2005, \mn@doi [\ssr] {10.1007/s11214-005-5097-2},
  \href {https://ui.adsabs.harvard.edu/abs/2005SSRv..120..165B} {120, 165}

\bibitem[\protect\citeauthoryear{{Caratti o Garatti}, {Giannini}, {Lorenzetti},
  {Massi}, {Nisini}  \& {Vitali}}{{Caratti o Garatti}
  et~al.}{2004}]{2004A&A...422..141C}
{Caratti o Garatti} A.,  {Giannini} T.,  {Lorenzetti} D.,  {Massi} F.,
  {Nisini} B.,   {Vitali} F.,  2004, \mn@doi [\aap]
  {10.1051/0004-6361:20047053}, \href
  {https://ui.adsabs.harvard.edu/abs/2004A&A...422..141C} {422, 141}

\bibitem[\protect\citeauthoryear{{Celli}, {Morlino}, {Gabici}  \&
  {Aharonian}}{{Celli} et~al.}{2019}]{2019MNRAS.490.4317C}
{Celli} S.,  {Morlino} G.,  {Gabici} S.,   {Aharonian} F.~A.,  2019, \mn@doi
  [\mnras] {10.1093/mnras/stz2897}, \href
  {https://ui.adsabs.harvard.edu/abs/2019MNRAS.490.4317C} {490, 4317}

\bibitem[\protect\citeauthoryear{{Dickey} \& {Lockman}}{{Dickey} \&
  {Lockman}}{1990}]{1990ARA&A..28..215D}
{Dickey} J.~M.,  {Lockman} F.~J.,  1990, \mn@doi [\araa]
  {10.1146/annurev.aa.28.090190.001243}, \href
  {https://ui.adsabs.harvard.edu/abs/1990ARA&A..28..215D} {28, 215}

\bibitem[\protect\citeauthoryear{{Drury}}{{Drury}}{1983}]{1983RPPh...46..973D}
{Drury} L.~O.,  1983, \mn@doi [Reports on Progress in Physics]
  {10.1088/0034-4885/46/8/002}, \href
  {https://ui.adsabs.harvard.edu/abs/1983RPPh...46..973D} {46, 973}

\bibitem[\protect\citeauthoryear{{Dubner} \& {Arnal}}{{Dubner} \&
  {Arnal}}{1988}]{1988A&AS...75..363D}
{Dubner} G.~M.,  {Arnal} E.~M.,  1988, \aaps, \href
  {https://ui.adsabs.harvard.edu/abs/1988A&AS...75..363D} {75, 363}

\bibitem[\protect\citeauthoryear{{Fontani}, {Beltr{\'a}n}, {Brand}, {Cesaroni},
  {Testi}, {Molinari}  \& {Walmsley}}{{Fontani}
  et~al.}{2005}]{2005A&A...432..921F}
{Fontani} F.,  {Beltr{\'a}n} M.~T.,  {Brand} J.,  {Cesaroni} R.,  {Testi} L.,
  {Molinari} S.,   {Walmsley} C.~M.,  2005, \mn@doi [\aap]
  {10.1051/0004-6361:20041810}, \href
  {https://ui.adsabs.harvard.edu/abs/2005A&A...432..921F} {432, 921}

\bibitem[\protect\citeauthoryear{{Fujita}, {Takahara}, {Ohira}  \&
  {Iwasaki}}{{Fujita} et~al.}{2011}]{2011MNRAS.415.3434F}
{Fujita} Y.,  {Takahara} F.,  {Ohira} Y.,   {Iwasaki} K.,  2011, \mn@doi
  [\mnras] {10.1111/j.1365-2966.2011.18980.x}, \href
  {https://ui.adsabs.harvard.edu/abs/2011MNRAS.415.3434F} {415, 3434}

\bibitem[\protect\citeauthoryear{Gabici \& Aharonian}{Gabici \&
  Aharonian}{2007}]{Gabici_2007}
Gabici S.,  Aharonian F.~A.,  2007, \mn@doi [The Astrophysical Journal]
  {10.1086/521047}, 665, L131

\bibitem[\protect\citeauthoryear{{Gabici}, {Aharonian}  \& {Casanova}}{{Gabici}
  et~al.}{2009}]{2009MNRAS.396.1629G}
{Gabici} S.,  {Aharonian} F.~A.,   {Casanova} S.,  2009, \mn@doi [\mnras]
  {10.1111/j.1365-2966.2009.14832.x}, \href
  {https://ui.adsabs.harvard.edu/abs/2009MNRAS.396.1629G} {396, 1629}

\bibitem[\protect\citeauthoryear{{Garay}, {Brooks}, {Mardones}  \&
  {Norris}}{{Garay} et~al.}{2003}]{2003ApJ...587..739G}
{Garay} G.,  {Brooks} K.~J.,  {Mardones} D.,   {Norris} R.~P.,  2003, \mn@doi
  [\apj] {10.1086/368310}, \href
  {https://ui.adsabs.harvard.edu/abs/2003ApJ...587..739G} {587, 739}

\bibitem[\protect\citeauthoryear{{Graham}}{{Graham}}{1991}]{1991PASP..103...79G}
{Graham} J.~A.,  1991, \mn@doi [\pasp] {10.1086/132797}, \href
  {https://ui.adsabs.harvard.edu/abs/1991PASP..103...79G} {103, 79}

\bibitem[\protect\citeauthoryear{{H.~E.~S.~S. Collaboration}
  et~al.,}{{H.~E.~S.~S. Collaboration} et~al.}{2015}]{2015A&A...575A..81H}
{H.~E.~S.~S. Collaboration} et~al., 2015, \mn@doi [\aap]
  {10.1051/0004-6361/201424805}, \href
  {https://ui.adsabs.harvard.edu/abs/2015A&A...575A..81H} {575, A81}

\bibitem[\protect\citeauthoryear{{Hanabata} et~al.,}{{Hanabata}
  et~al.}{2014}]{2014ApJ...786..145H}
{Hanabata} Y.,  et~al., 2014, \mn@doi [\apj] {10.1088/0004-637X/786/2/145},
  \href {https://ui.adsabs.harvard.edu/abs/2014ApJ...786..145H} {786, 145}

\bibitem[\protect\citeauthoryear{Hewitt, Grondin, Lemoine-Goumard, Reposeur,
  Ballet  \& Tanaka}{Hewitt et~al.}{2012}]{Hewitt_2012}
Hewitt J.~W.,  Grondin M.-H.,  Lemoine-Goumard M.,  Reposeur T.,  Ballet J.,
  Tanaka T.,  2012, \mn@doi [The Astrophysical Journal]
  {10.1088/0004-637x/759/2/89}, 759, 89

\bibitem[\protect\citeauthoryear{{Kafexhiu}, {Aharonian}, {Taylor}  \&
  {Vila}}{{Kafexhiu} et~al.}{2014}]{2014PhRvD..90l3014K}
{Kafexhiu} E.,  {Aharonian} F.,  {Taylor} A.~M.,   {Vila} G.~S.,  2014, \mn@doi
  [\prd] {10.1103/PhysRevD.90.123014}, \href
  {https://ui.adsabs.harvard.edu/abs/2014PhRvD..90l3014K} {90, 123014}

\bibitem[\protect\citeauthoryear{{Kalberla}, {Burton}, {Hartmann}, {Arnal},
  {Bajaja}, {Morras}  \& {P{\"o}ppel}}{{Kalberla}
  et~al.}{2005}]{2005A&A...440..775K}
{Kalberla} P.~M.~W.,  {Burton} W.~B.,  {Hartmann} D.,  {Arnal} E.~M.,  {Bajaja}
  E.,  {Morras} R.,   {P{\"o}ppel} W.~G.~L.,  2005, \mn@doi [\aap]
  {10.1051/0004-6361:20041864}, \href
  {https://ui.adsabs.harvard.edu/abs/2005A&A...440..775K} {440, 775}

\bibitem[\protect\citeauthoryear{{Kerby}, {Kaur}, {Falcone}, {Stroh},
  {Ferrara}, {Kennea}  \& {Colosimo}}{{Kerby}
  et~al.}{2021}]{2021AJ....161..154K}
{Kerby} S.,  {Kaur} A.,  {Falcone} A.~D.,  {Stroh} M.~C.,  {Ferrara} E.~C.,
  {Kennea} J.~A.,   {Colosimo} J.,  2021, \mn@doi [\aj]
  {10.3847/1538-3881/abda53}, \href
  {https://ui.adsabs.harvard.edu/abs/2021AJ....161..154K} {161, 154}

\bibitem[\protect\citeauthoryear{{Lande} et~al.,}{{Lande}
  et~al.}{2012a}]{2012ApJ...756....5L}
{Lande} J.,  et~al., 2012a, \mn@doi [\apj] {10.1088/0004-637X/756/1/5}, \href
  {https://ui.adsabs.harvard.edu/abs/2012ApJ...756....5L} {756, 5}

\bibitem[\protect\citeauthoryear{Lande et~al.,}{Lande
  et~al.}{2012b}]{Lande_2012}
Lande J.,  et~al., 2012b, \mn@doi [The Astrophysical Journal]
  {10.1088/0004-637x/756/1/5}, 756, 5

\bibitem[\protect\citeauthoryear{{Li} \& {Chen}}{{Li} \&
  {Chen}}{2010}]{2010MNRAS.409L..35L}
{Li} H.,  {Chen} Y.,  2010, \mn@doi [\mnras]
  {10.1111/j.1745-3933.2010.00944.x}, \href
  {https://ui.adsabs.harvard.edu/abs/2010MNRAS.409L..35L} {409, L35}

\bibitem[\protect\citeauthoryear{{Lorenzetti}, {Giannini}, {Vitali}, {Massi}
  \& {Nisini}}{{Lorenzetti} et~al.}{2002}]{2002ApJ...564..839L}
{Lorenzetti} D.,  {Giannini} T.,  {Vitali} F.,  {Massi} F.,   {Nisini} B.,
  2002, \mn@doi [\apj] {10.1086/324305}, \href
  {https://ui.adsabs.harvard.edu/abs/2002ApJ...564..839L} {564, 839}

\bibitem[\protect\citeauthoryear{{MAGIC Collaboration} et~al.,}{{MAGIC
  Collaboration} et~al.}{2020}]{2020arXiv201015854M}
{MAGIC Collaboration} et~al., 2020, arXiv e-prints, \href
  {https://ui.adsabs.harvard.edu/abs/2020arXiv201015854M} {p. arXiv:2010.15854}

\bibitem[\protect\citeauthoryear{{Malkov}, {Diamond}  \& {Sagdeev}}{{Malkov}
  et~al.}{2011}]{2011NatCo...2..194M}
{Malkov} M.~A.,  {Diamond} P.~H.,   {Sagdeev} R.~Z.,  2011, \mn@doi [Nature
  Communications] {10.1038/ncomms1195}, \href
  {https://ui.adsabs.harvard.edu/abs/2011NatCo...2..194M} {2, 194}

\bibitem[\protect\citeauthoryear{{Malkov}, {Diamond}, {Sagdeev}, {Aharonian}
  \& {Moskalenko}}{{Malkov} et~al.}{2013}]{2013ApJ...768...73M}
{Malkov} M.~A.,  {Diamond} P.~H.,  {Sagdeev} R.~Z.,  {Aharonian} F.~A.,
  {Moskalenko} I.~V.,  2013, \mn@doi [\apj] {10.1088/0004-637X/768/1/73}, \href
  {https://ui.adsabs.harvard.edu/abs/2013ApJ...768...73M} {768, 73}

\bibitem[\protect\citeauthoryear{{Mattox} et~al.,}{{Mattox}
  et~al.}{1996}]{1996ApJ...461..396M}
{Mattox} J.~R.,  et~al., 1996, \mn@doi [\apj] {10.1086/177068}, \href
  {https://ui.adsabs.harvard.edu/abs/1996ApJ...461..396M} {461, 396}

\bibitem[\protect\citeauthoryear{{Mauch}, {Murphy}, {Buttery}, {Curran},
  {Hunstead}, {Piestrzynski}, {Robertson}  \& {Sadler}}{{Mauch}
  et~al.}{2003}]{2003MNRAS.342.1117M}
{Mauch} T.,  {Murphy} T.,  {Buttery} H.~J.,  {Curran} J.,  {Hunstead} R.~W.,
  {Piestrzynski} B.,  {Robertson} J.~G.,   {Sadler} E.~M.,  2003, \mn@doi
  [\mnras] {10.1046/j.1365-8711.2003.06605.x}, \href
  {https://ui.adsabs.harvard.edu/abs/2003MNRAS.342.1117M} {342, 1117}

\bibitem[\protect\citeauthoryear{{Mitchell}, {Rowell}, {Celli}  \&
  {Einecke}}{{Mitchell} et~al.}{2021}]{2021MNRAS.503.3522M}
{Mitchell} A.~M.~W.,  {Rowell} G.~P.,  {Celli} S.,   {Einecke} S.,  2021,
  \mn@doi [\mnras] {10.1093/mnras/stab667}, \href
  {https://ui.adsabs.harvard.edu/abs/2021MNRAS.503.3522M} {503, 3522}

\bibitem[\protect\citeauthoryear{{Mottram}, {Hoare}, {Lumsden}, {Oudmaijer},
  {Urquhart}, {Sheret}, {Clarke}  \& {Allsopp}}{{Mottram}
  et~al.}{2007}]{2007A&A...476.1019M}
{Mottram} J.~C.,  {Hoare} M.~G.,  {Lumsden} S.~L.,  {Oudmaijer} R.~D.,
  {Urquhart} J.~S.,  {Sheret} T.~L.,  {Clarke} A.~J.,   {Allsopp} J.,  2007,
  \mn@doi [\aap] {10.1051/0004-6361:20077663}, \href
  {https://ui.adsabs.harvard.edu/abs/2007A&A...476.1019M} {476, 1019}

\bibitem[\protect\citeauthoryear{{Nava}, {Gabici}, {Marcowith}, {Morlino}  \&
  {Ptuskin}}{{Nava} et~al.}{2016}]{2016MNRAS.461.3552N}
{Nava} L.,  {Gabici} S.,  {Marcowith} A.,  {Morlino} G.,   {Ptuskin} V.~S.,
  2016, \mn@doi [\mnras] {10.1093/mnras/stw1592}, \href
  {https://ui.adsabs.harvard.edu/abs/2016MNRAS.461.3552N} {461, 3552}

\bibitem[\protect\citeauthoryear{{Ohira} \& {Takahara}}{{Ohira} \&
  {Takahara}}{2010}]{2010ApJ...721L..43O}
{Ohira} Y.,  {Takahara} F.,  2010, \mn@doi [\apjl]
  {10.1088/2041-8205/721/1/L43}, \href
  {https://ui.adsabs.harvard.edu/abs/2010ApJ...721L..43O} {721, L43}

\bibitem[\protect\citeauthoryear{{Ohira}, {Murase}  \& {Yamazaki}}{{Ohira}
  et~al.}{2011}]{2011MNRAS.410.1577O}
{Ohira} Y.,  {Murase} K.,   {Yamazaki} R.,  2011, \mn@doi [\mnras]
  {10.1111/j.1365-2966.2010.17539.x}, \href
  {https://ui.adsabs.harvard.edu/abs/2011MNRAS.410.1577O} {410, 1577}

\bibitem[\protect\citeauthoryear{{Peron}, {Aharonian}, {Casanova}, {Zanin}  \&
  {Romoli}}{{Peron} et~al.}{2020}]{2020ApJ...896L..23P}
{Peron} G.,  {Aharonian} F.,  {Casanova} S.,  {Zanin} R.,   {Romoli} C.,  2020,
  \mn@doi [\apjl] {10.3847/2041-8213/ab93d1}, \href
  {https://ui.adsabs.harvard.edu/abs/2020ApJ...896L..23P} {896, L23}

\bibitem[\protect\citeauthoryear{{Ptuskin} \& {Zirakashvili}}{{Ptuskin} \&
  {Zirakashvili}}{2003}]{2003A&A...403....1P}
{Ptuskin} V.~S.,  {Zirakashvili} V.~N.,  2003, \mn@doi [\aap]
  {10.1051/0004-6361:20030323}, \href
  {https://ui.adsabs.harvard.edu/abs/2003A&A...403....1P} {403, 1}

\bibitem[\protect\citeauthoryear{Ptuskin, Moskalenko, Jones, Strong  \&
  Zirakashvili}{Ptuskin et~al.}{2006}]{Ptuskin_2006}
Ptuskin V.~S.,  Moskalenko I.~V.,  Jones F.~C.,  Strong A.~W.,   Zirakashvili
  V.~N.,  2006, \mn@doi [The Astrophysical Journal] {10.1086/501117}, 642, 902

\bibitem[\protect\citeauthoryear{{Recchia}, {Galli}, {Nava}, {Padovani},
  {Gabici}, {Marcowith}, {Ptuskin}  \& {Morlino}}{{Recchia}
  et~al.}{2021}]{2021arXiv210604948R}
{Recchia} S.,  {Galli} D.,  {Nava} L.,  {Padovani} M.,  {Gabici} S.,
  {Marcowith} A.,  {Ptuskin} V.,   {Morlino} G.,  2021, arXiv e-prints, \href
  {https://ui.adsabs.harvard.edu/abs/2021arXiv210604948R} {p. arXiv:2106.04948}

\bibitem[\protect\citeauthoryear{{Reynoso}, {Dubner}, {Goss}  \&
  {Arnal}}{{Reynoso} et~al.}{1995}]{1995AJ....110..318R}
{Reynoso} E.~M.,  {Dubner} G.~M.,  {Goss} W.~M.,   {Arnal} E.~M.,  1995,
  \mn@doi [\aj] {10.1086/117522}, \href
  {https://ui.adsabs.harvard.edu/abs/1995AJ....110..318R} {110, 318}

\bibitem[\protect\citeauthoryear{{Reynoso}, {Cichowolski}  \&
  {Walsh}}{{Reynoso} et~al.}{2017}]{2017MNRAS.464.3029R}
{Reynoso} E.~M.,  {Cichowolski} S.,   {Walsh} A.~J.,  2017, \mn@doi [\mnras]
  {10.1093/mnras/stw2219}, \href
  {https://ui.adsabs.harvard.edu/abs/2017MNRAS.464.3029R} {464, 3029}

\bibitem[\protect\citeauthoryear{{Reynoso}, {Vel{\'a}zquez}  \&
  {Cichowolski}}{{Reynoso} et~al.}{2018}]{2018MNRAS.477.2087R}
{Reynoso} E.~M.,  {Vel{\'a}zquez} P.~F.,   {Cichowolski} S.,  2018, \mn@doi
  [\mnras] {10.1093/mnras/sty751}, \href
  {https://ui.adsabs.harvard.edu/abs/2018MNRAS.477.2087R} {477, 2087}

\bibitem[\protect\citeauthoryear{{Schlafly} \& {Finkbeiner}}{{Schlafly} \&
  {Finkbeiner}}{2011}]{2011ApJ...737..103S}
{Schlafly} E.~F.,  {Finkbeiner} D.~P.,  2011, \mn@doi [\apj]
  {10.1088/0004-637X/737/2/103}, \href
  {https://ui.adsabs.harvard.edu/abs/2011ApJ...737..103S} {737, 103}

\bibitem[\protect\citeauthoryear{{Suzuki}, {Bamba}, {Yamazaki}  \&
  {Ohira}}{{Suzuki} et~al.}{2020}]{2020PASJ...72...72S}
{Suzuki} H.,  {Bamba} A.,  {Yamazaki} R.,   {Ohira} Y.,  2020, \mn@doi [\pasj]
  {10.1093/pasj/psaa061}, \href
  {https://ui.adsabs.harvard.edu/abs/2020PASJ...72...72S} {72, 72}

\bibitem[\protect\citeauthoryear{{Uchiyama}, {Funk}, {Katagiri}, {Katsuta},
  {Lemoine-Goumard}, {Tajima}, {Tanaka}  \& {Torres}}{{Uchiyama}
  et~al.}{2012}]{2012ApJ...749L..35U}
{Uchiyama} Y.,  {Funk} S.,  {Katagiri} H.,  {Katsuta} J.,  {Lemoine-Goumard}
  M.,  {Tajima} H.,  {Tanaka} T.,   {Torres} D.~F.,  2012, \mn@doi [\apjl]
  {10.1088/2041-8205/749/2/L35}, \href
  {https://ui.adsabs.harvard.edu/abs/2012ApJ...749L..35U} {749, L35}

\bibitem[\protect\citeauthoryear{{Urquhart} et~al.,}{{Urquhart}
  et~al.}{2007}]{2007A&A...474..891U}
{Urquhart} J.~S.,  et~al., 2007, \mn@doi [\aap] {10.1051/0004-6361:20078025},
  \href {https://ui.adsabs.harvard.edu/abs/2007A&A...474..891U} {474, 891}

\bibitem[\protect\citeauthoryear{{Wilms}, {Allen}  \& {McCray}}{{Wilms}
  et~al.}{2000}]{2000ApJ...542..914W}
{Wilms} J.,  {Allen} A.,   {McCray} R.,  2000, \mn@doi [\apj] {10.1086/317016},
  \href {https://ui.adsabs.harvard.edu/abs/2000ApJ...542..914W} {542, 914}

\bibitem[\protect\citeauthoryear{{Winkler} \& {Kirshner}}{{Winkler} \&
  {Kirshner}}{1985}]{1985ApJ...299..981W}
{Winkler} P.~F.,  {Kirshner} R.~P.,  1985, \mn@doi [\apj] {10.1086/163764},
  \href {https://ui.adsabs.harvard.edu/abs/1985ApJ...299..981W} {299, 981}

\bibitem[\protect\citeauthoryear{{Woermann}, {Gaylard}  \&
  {Otrupcek}}{{Woermann} et~al.}{2000}]{2000MNRAS.317..421W}
{Woermann} B.,  {Gaylard} M.~J.,   {Otrupcek} R.,  2000, \mn@doi [\mnras]
  {10.1046/j.1365-8711.2000.03575.x}, \href
  {https://ui.adsabs.harvard.edu/abs/2000MNRAS.317..421W} {317, 421}

\bibitem[\protect\citeauthoryear{{Xin}, {Guo}, {Liao}, {Yuan}, {Liu}  \&
  {Wei}}{{Xin} et~al.}{2017}]{2017ApJ...843...90X}
{Xin} Y.-L.,  {Guo} X.-L.,  {Liao} N.-H.,  {Yuan} Q.,  {Liu} S.-M.,   {Wei}
  D.-M.,  2017, \mn@doi [\apj] {10.3847/1538-4357/aa74bb}, \href
  {https://ui.adsabs.harvard.edu/abs/2017ApJ...843...90X} {843, 90}

\bibitem[\protect\citeauthoryear{{Zabalza}}{{Zabalza}}{2015}]{naima}
{Zabalza} V.,  2015, \mn@doi [Proceedings of The 34th International Cosmic Ray
  Conference, The Hague, The Netherlands] {10.22323/1.236.0922}, \href
  {http://adsabs.harvard.edu/abs/2015arXiv150903319Z} {236, 922}

\makeatother
\end{thebibliography}

% Don't change these lines
\bsp    % typesetting comment
\label{lastpage}

\end{document}